\UseRawInputEncoding
\documentclass[twoside]{article}
\pdfoutput=1
\usepackage{amsfonts,amssymb,amsbsy,textcomp,marvosym,picins,amsmath,caption,threeparttable,amsthm,subfigure,float,lastpage,lscape,color}
\usepackage{eurosym,mathrsfs,fancyhdr,CJK,multicol,graphics,indentfirst,color,bm,upgreek,booktabs,graphicx,multirow,warpcol}
\usepackage{epstopdf}
\usepackage{cite}
\usepackage{url}
\usepackage{subfigure}

\def\footnoterule{\kern 1mm \hrule width 10cm \kern 2mm}

\allowdisplaybreaks
\catcode`@=11
\def\title#1{\vspace{3mm}\begin{flushleft}\vglue-.1cm\Large\bf\boldmath\protect\baselineskip=18pt plus.2pt minus.1pt #1
\end{flushleft}\vspace{1mm} }
\def\author#1{\begin{flushleft}\normalsize #1\end{flushleft}\vspace*{-4pt} \vspace{3mm}}
\def\address#1#2{\begin{flushleft}\vglue-.35cm${}^{#1}$\small\it #2\vglue-.35cm\end{flushleft}\vspace{-2mm}\par}
\def\jz#1#2{{$^{\footnotesize\textcircled{\tiny #1}}$\let\thefootnote\relax\footnotetext{\!\!$^{\footnotesize\textcircled{\tiny #1}}$#2}}}
\catcode`@=11
\def\section{\@startsection{section}{1}{\z@}%
 {-3ex \@plus -.3ex \@minus -.2ex}%
 {2.2ex \@plus.2ex}%
{\normalfont\normalsize\protect\baselineskip=14.5pt plus.2pt minus.2pt\bfseries}}
\def\subsection{\@startsection{subsection}{2}{\z@}%
 {-3ex\@plus -.2ex \@minus -.2ex}%
 {2ex \@plus.2ex}%
{\normalfont\normalsize\protect\baselineskip=12.5pt plus.2pt minus.2pt\bfseries}}
\def\subsubsection{\@startsection{subsubsection}{3}{\z@}%
 {-2.2ex\@plus -.21ex \@minus -.2ex}%
 {1.4ex \@plus.2ex}
{\normalfont\normalsize\protect\baselineskip=12pt plus.2pt minus.2pt\sl}}

\begin{document}
\begin{CJK*}{GBK}{song}
\thispagestyle{empty}
\vspace*{-13mm}
\noindent {\small Journal of computer science and technology: Instruction for authors.
JOURNAL OF COMPUTER SCIENCE AND TECHNOLOGY}
\vspace*{2mm}

\title{ISC4DGF: Enhancing Directed Grey-box Fuzzing with LLM-Driven Initial Seed Corpus Generation}

\author{Yijiang Xu$^{1}$, Hongrui Jia$^{1}$, Liguo Chen$^{1}$, Xin Wang$^{1}$, Zhengran Zeng$^{1}$, Yidong Wang$^{1}$, Qing Gao$^{1}$, Jindong Wang$^{2}$, Wei Ye$^{1}$, Shikun Zhang$^{1}$$^{*}$, Zhonghai Wu$^{1}$}
\address{1}{Peking University, Beijing, China}
\address{2}{Microsoft Research Asia, Beijing, China}

\let\thefootnote\relax\footnotetext{{}\\[-4mm]\indent\ Regular Paper}
\let\thefootnote\relax\footnotetext{{}\\[-4mm]\indent\ * corresponding authors.}

\noindent {\small\bf Abstract} \quad  {\small { Fuzz testing is crucial for identifying software vulnerabilities, with coverage-guided grey-box fuzzers like AFL and Angora excelling in broad detection. However, as the need for targeted detection grows, directed grey-box fuzzing (DGF) has become essential, focusing on specific vulnerabilities. The initial seed corpus, which consists of carefully selected input samples that the fuzzer uses as a starting point, is fundamental in determining the paths that the fuzzer explores. A well-designed seed corpus can guide the fuzzer more effectively towards critical areas of the code, improving the efficiency and success of the fuzzing process. Even with its importance, many works concentrate on refining guidance mechanisms while paying less attention to optimizing the initial seed corpus.

In this paper, we introduce ISC4DGF, a novel approach to generating optimized initial seed corpus for DGF using Large Language Models (LLMs). By leveraging LLMs' deep software understanding and refined user inputs, ISC4DGF creates precise seed corpus that efficiently trigger specific vulnerabilities. Implemented on AFL and tested against state-of-the-art fuzzers like AFLGo, FairFuzz, and Entropic using the Magma benchmark, ISC4DGF achieved a 35.63x speedup and 616.10x fewer target reaches. Moreover, ISC4DGF focused on more effectively detecting target vulnerabilities, enhancing efficiency while operating with reduced code coverage. }}

\vspace*{3mm}


\vspace*{4mm}

\end{CJK*}
\baselineskip=18pt plus.2pt minus.2pt
\parskip=0pt plus.2pt minus0.2pt
\begin{multicols}{2}

\section{Introduction}

Fuzz testing \cite{sutton2007fuzzing} is a crucial technique for software security. By providing a program with a wide range of inputs, it can trigger unexpected behaviors or crashes, helping developers identify and fix vulnerabilities early, which reduces the cost of later-stage fixes. In both research and industry, fuzz testing has proven to be an effective way to make software more secure.

Over the years, various methods have been developed to improve the effectiveness of fuzz testing. Coverage-guided grey-box fuzzing, such as AFL \cite{afl} and Angora \cite{chen2018angora}, focuses on increasing the amount of code that gets tested, which improves the chances of finding unknown vulnerabilities. Directed grey-box fuzzing (DGF), such as AFLGo \cite{aflgo} and BEACON \cite{beacon}, goes a step further by focusing on specific vulnerabilities. These methods use strategies including heuristic guidance strategies and seed mutation techniques to enhance fuzzing efficiency, especially in scenarios like patch testing \cite{bohme2013regression}, crash reproduction \cite{pham2015hercules}, and verifying static analysis reports \cite{christakis2016guiding}.

However, DGF methods still face challenges, particularly in the selection and optimization of the initial set of test cases, known as the seed corpus. The success of DGF depends heavily on the effectiveness of its guidance mechanisms, yet the role of the initial seed corpus is often underestimated. Research \cite{rebert2014optimizing} indicates that the quality of this initial seed corpus is crucial for the success of fuzzing. Herrera et al. \cite{herrera2021seed} found that different initial seed selections can significantly influence the outcomes, with well-chosen seeds greatly enhancing the efficiency of vulnerability detection. Nevertheless, few studies have addressed the design and optimization of the initial seed corpus for directed fuzzers, despite researchers frequently recognizing the importance of high-quality input seed corpus and its impact on fuzzer performance \cite{rebert2014optimizing, huang2023titan}.

Large Language Models (LLMs) have recently made impressive advances, particularly in their ability to understand and generate code. Trained on vast datasets, LLMs excel at recognizing logical structures, identifying patterns in software, and comprehending complex programming languages, making them powerful tools beyond natural language processing.
Given their strong program understanding capabilities, LLMs are well-suited for generating the initial seed corpus in DGF. While some studies \cite{deng2023large, deng2023large2, xia2024fuzz4all} have combined fuzzing with LLMs, they focused on using prompts to guide seed mutations, demonstrating the potential of LLMs to produce effective inputs for coverage-based fuzzers. However, there has been little exploration into using prompts to generate the initial seed corpus specifically tailored for DGF.

To overcome the above challenges, we propose ISC4DGF, an effective initial seed corpus generation approach for DGF based on LLM. ISC4DGF leverages the advanced program comprehension capabilities of LLMs, guided by essential user-provided details through prompts. This approach enables ISC4DGF to produce a highly focused and optimized seed set tailored to trigger specific vulnerabilities during fuzz testing.
The ISC4DGF process begins by receiving detailed input from the user, which include project introduction, driver source code, CVE details, and CVE corresponding patches. Recognizing that these inputs can be lengthy, redundant, or even partially irrelevant, ISC4DGF first employs a refinement LLM to optimize this information into concise and informative prompts.
Once the user inputs have been refined, these optimized prompts are fed into generation LLM, which is tasked with generating a diverse set of potential seed inputs. These seeds are specifically designed to explore critical paths within the code and to trigger the vulnerabilities identified in the user input. The generated seeds are then evaluated, scored, and ranked based on their effectiveness. ISC4DGF selects the best candidate prompts from this pool to form the initial seed corpus for the fuzzing process.
After generating the optimized seed corpus, ISC4DGF replaces the default seed corpus provided by the fuzzer with these new seeds. Then, the fuzzer is executed, using the optimized seeds, and continues to run until it either discovers the target vulnerability or the testing period concludes. 

We implemented ISC4DGF on AFL 2.57b\cite{afl} and compared it with the state-of-the-art directed fuzzer AFLGo \cite{aflgo} as well as widely-used coverge-based fuzzers AFL \cite{afl}, FairFuzz \cite{fairfuzz}, and Entropic \cite{entropic}. The comparison was conducted on real-world programs and the advanced fuzzing benchmark Magma \cite{magma}. On average, ISC4DGF achieved a speedup of 35.63x and triggered the target with 616.10x fewer target reaches. Additionally, ISC4DGF effectively narrowed the execution scope of the fuzzing process, achieving 8.12x more target reaches with a smaller code coverage compared to the other fuzzers.

To sum up, we make the following contributions:

\begin{itemize}
    \item We propose ISC4DGF, the first effective DGF initial seed corpus generation method based on LLM prompts.
    
    \item We introduce a prompt refinement strategy and a method for selecting LLM-generated seeds, effectively enabling the extraction of relevant user input and the selection of high-quality seeds.
    
    \item We evaluate ISC4DGF on the widely-used Magma dataset. ISC4DGF is able to trigger specified bugs faster than state-of-the-art fuzzers, while also focusing the fuzzer's execution more closely on the target.
\end{itemize}

\section{Background}
\subsection{Large Language Models}

Pre-trained Large Language Models (LLMs) are advanced neural networks with billions of parameters. These models are trained on large amounts of text data using an autoregressive method, where the model learns to predict the next word in a sequence based on the previous context. This extensive training allows LLMs to act as one-shot or even zero-shot learners \cite{mann2020language}, enabling them to perform a wide variety of tasks with little or no additional training.

LLMs are typically used for specific tasks through either fine-tuning \cite{radford2018improving} or prompting \cite{liu2023pre}. Fine-tuning involves further training the model on a task-specific dataset, which adjusts its internal weights to enhance performance for that task. However, this approach can be challenging due to the potential lack of suitable datasets, and as LLMs grow in size \cite{kaplan2020scaling}, the cost and complexity of fine-tuning also increase.
Alternatively, prompting enables the LLM to perform tasks without altering its weights. The process involves providing the model with a detailed description of the task, sometimes including a few examples to illustrate how to solve it. By utilizing the model's existing knowledge, users can guide its performance through a technique known as prompt engineering \cite{liu2023pre}, where different input instructions are tested to identify the most effective prompts.

\subsection{Fuzzing and Testing}

Fuzz testing is a important technique for improving software security. The main idea is to provide the target program with a large number of varied inputs as test cases, which helps expose software vulnerabilities by triggering exceptions. Developers can then identify and fix these vulnerabilities earlier, reducing the cost of later repairs.

Fuzzer test case generation methods are typically divided into generation-based and mutation-based approaches. Generation-based fuzzers, such as QuickFuzz \cite{quickfuzz}, Gandalf \cite{Gandalf}, and CodeAlchemist \cite{codealchemist}, generate test cases based on the input format specifications of the program being tested. In contrast, mutation-based fuzzers, like AFL \cite{afl}, libFuzzer \cite{libFuzzer}, AFLGo \cite{aflgo}, and Beacon \cite{beacon}, start with an initial set of seed inputs and create new test cases by modifying these seeds.

American Fuzzy Lop (AFL) \cite{afl}, one of the most popular coverage-guided grey-box fuzzers, uses lightweight code instrumentation to collect coverage data during fuzzing. Feeding this coverage information back into the mutation algorithm helps guide the fuzzer to explore new code paths in the target program. 
While coverage-guided fuzzers like AFL excel at discovering a broad range of vulnerabilities by exploring as much of the program's code as possible, they are not specifically optimized for targeting particular vulnerabilities. Directed grey-box fuzzers are designed to focus the fuzzing process on specific areas of the code that are more likely to contain known or suspected vulnerabilities. Instead of spreading resources across the entire codebase, directed fuzzers concentrate on reaching and testing specific targets more efficiently.
AFLGo \cite{aflgo}, the first directed grey-box fuzzer, introduced the concept of targeting specific vulnerabilities within coverage-guided grey-box fuzz testing. By employing a heuristic distance algorithm based on static analysis, AFLGo measures the proximity between seeds and test targets. Prioritizing seeds that are closer to the target allows the fuzzer to trigger specific vulnerabilities more quickly.

\section{Motivation}

\begin{table*}
  \centering
  \caption{\textbf{Bug-finding results.} \textbf{BUG ID} comes from Magma, \textbf{CVE} is the CVE corresponding to the bug, \textbf{trigger} is the time required for the fuzzer to trigger the bug (in seconds), and \textbf{reach} is the number of times the fuzzer reaches the bug.}
  \begin{tabular}{cccccccc}
    \toprule
    \multirow{2}{*}{\textbf{Bug ID}} & \multirow{2}{*}{\textbf{CVE}} & \multicolumn{2}{c}{\textbf{AFL}} & \multicolumn{2}{c}{\textbf{AFL-random}}\\
    \cline{3-4}
    \cline{5-6}
    \textbf{}&\textbf{}&\textbf{trigger}&\textbf{reach}&\textbf{trigger}&\textbf{reach}\\
    \midrule
    \ PNG001 & CVE-2018-13785 & T.O & 28333100 & T.O.& 15817402\\
    \ PNG003 & CVE-2015-8472 & 15 & 58190 & T.O.& 0\\
    \ PNG007 & CVE-2013-6954 & T.O & 10951093 & T.O.& 0\\
    \ TIF007 & CVE-2016-10270 & 6812 & 1880 & 24 & 2172\\
    \ TIF012 & CVE-2016-3658 & 14058 & 28300878 & T.O. & 14901472\\
    \ TIF014 & CVE-2017-11613 & 18505 & 363487 & 24920 & 5492969\\
    \midrule
    \ \textbf{Avg.} &	- & - & \textbf{11334771} & - &\textbf{6035669}\\
  \bottomrule
\end{tabular}
\label{random}
\end{table*}

Mutation-based grey-box fuzzing, the most widely used fuzzing technique, typically relies on a set of non-crashing initial seed inputs to guide the bug-finding process. However, Klees et al. \cite{klees2018evaluating} found that many studies assume all seeds are equally effective, often neglecting the optimization or careful selection of initial seeds. They noted that "a fuzzer's performance can vary significantly depending on the seed used" and recommended that researchers be explicit about their seed selection methods.

Herrera et al. \cite{herrera2021seed} reviewed 28 more studies and categorized initial seed selection into four types: benchmark and fuzzer-provided seeds, manually-constructed seeds, random seeds, and empty seeds. Many works, such as Hawkeye \cite{hawkeye} and FuzzFactory \cite{padhye2019fuzzfactory}, used fuzzers provided seeds without specifying details. Redqueen \cite{aschermann2019redqueen} and Grimoire \cite{blazytko2019grimoire}, manually created generic seeds but didn't explain their choices or the impact on testing outcomes. Studies using random or empty seeds \cite{lyu2019mopt, bohme2020boosting} also lacked clarity on how initial seeds influence fuzz testing results.

To highlight the importance of initial seed selection, we conducted a series of comparative experiments using the Magma dataset \cite{magma}. Magma comprises open-source libraries with extensive usage and a history of security-critical bugs and vulnerabilities, including front-ported bugs from previous reports to the latest library versions \cite{magma-website}. Magma provides a set of available and efficient initial seed sets for each fuzz test object. In our experiments, we used the prompt instruction: "Please generate a random $\langle format \rangle$ test case" to randomly generate 10 initial seeds with GPT-4-0613, replacing the initial seed set provided by Magma. We then conducted a 24-hour experimental comparison. The results are presented in Table \ref{random}.

\begin{figure*}
    \centering
    \includegraphics[width=0.92\linewidth]{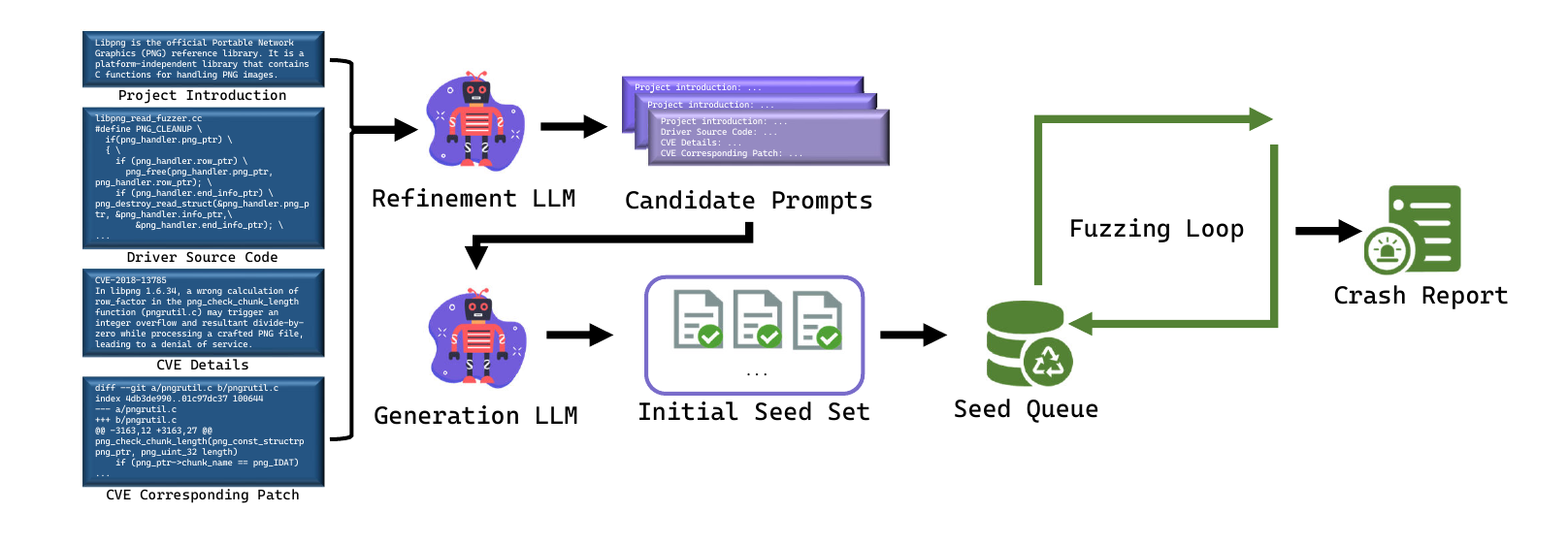}
    \caption{\textbf{Overview of ISC4DGF.} Illustrates the process of refining user inputs through a refinement LLM, followed by the use of a generation LLM to create and select the optimized initial seed corpus for directed grey-box fuzzing.}
    \label{overview}
\end{figure*}

We used AFL as the benchmark fuzzer and measured the time required to trigger the test target using different initial seeds, aiming to assess the impact of the initial seed corpus on fuzz testing efficiency and to count the number of target reaches, thereby evaluating the seed's ability to trigger the target. Compared to fuzzers using randomly generated initial seed corpus, the corpus provided by Magma achieved faster target triggering with more reaches. Notably, in our experiment, AFL with randomly generated seeds failed to reach PNG003 and PNG007 within 24 hours. This demonstrates that a well-designed initial seed corpus can enhance fuzz testing efficiency.
Interestingly, TIF007 achieved faster target triggering with randomly generated initial seed corpus, underscoring the importance of initial seed corpus design. Even random generation can improve test efficiency in some cases, suggesting that a more effective seed design approach could further optimize the fuzzer's performance.

\section{Approach}

In this work, we propose ISC4DGF, an LLM-based approach for efficient initial seed corpus generation for DGF. ISC4DGF leverages the deep program understanding capabilities of LLMs and combines them with critical user-provided information to generate an optimized corpus of initial seeds that can effectively trigger the specified target. Figure \ref{overview} provides an overview of our approach.

\subsection{Overview}

ISC4DGF enhances the fuzzing process by using a refinement LLM to optimize user inputs and a generation LLM to produce and select targeted initial seed inputs.
ISC4DGF begins by receiving relevant user input, including project details, driver source code, CVE information, and CVE corresponding patches. Since user input may be lengthy, redundant, or partially irrelevant, we utilize a large, state-of-the-art LLM to refine these prompts into a concise and informative form. ISC4DGF then feeds the optimized prompts into the generation LLM, which scores and ranks the generated results, selecting the most suitable candidates as the initial seed corpus.
The original seed set provided by the fuzzer is replaced with the optimized seed set generated by the LLM. The fuzzer is then executed until it either identifies the specified bug or the testing period ends. Crash reports generated during this process are used to analyze and validate the capabilities of ISC4DGF.

\subsection{Refining Prompts}

The following section provides a detailed introduction to the first of the two main steps in ISC4DGF, which involves refining and extracting user input from a given template through the refinement LLM to generate prompts suitable for directed fuzzing. The focus of ISC4DGF is on selecting an appropriate initial seed corpus to enable efficient exploration during directed fuzzing, ultimately discovering the unique target specified by the user. Therefore, we consider user input that describes both the general context of the program under test (PUT) and the specifics of the target vulnerability. User input specifically includes project introduction, driver source code, CVE details, and the CVE corresponding patch.

\begin{itemize}
    \item \textbf{Project introduction:} Provides an overview of the PUT's functionality, enhancing the LLM's comprehension of the program.

    \item \textbf{Driver source code:} Ensures that the generated seeds conform to the expected format, as non-compliant seeds may lead to ineffective or inefficient fuzzing.

    \item \textbf{CVE details:} Specifies a clear target for seed generation, including key information such as CVE numbers, descriptions, and vulnerability types.

    \item \textbf{CVE corresponding patch:} Improves the specificity of seed generation by leveraging detailed information from the vulnerability patch, enhancing the LLM's understanding of the specific vulnerability.
 
\end{itemize}

Unlike traditional fuzzers that require input in a specific format, ISC4DGF can directly interpret natural language descriptions or code examples provided by the user. The four types of input in ISC4DGF are manually specified by the user and may include some redundant or irrelevant information. Consequently, using these inputs directly as prompts for generating LLM outputs may not be effective. To address this, ISC4DGF employs a refinement process where the refinement LLM prunes and optimizes the user-provided prompts to generate higher-quality inputs, thereby enabling more effective LLM-based fuzz testing.

We use the prompt: "Please summarize the above information concisely to describe the functionality of the test project, the usage of the fuzz driver, and the details of the vulnerability." ISC4DGF first generates an initial candidate prompt using greedy sampling with a temperature of 0, enabling the refinement LLM to produce a reliable solution with high confidence. This approach has been widely applied in other fields, such as program synthesis \cite{chen2021evaluating}. Following this, ISC4DGF continues to sample at a higher temperature to generate more diverse prompts, similar to methods used in previous studies \cite{chen2021evaluating,xia2024fuzz4all}. High-temperature sampling generates different prompts, each offering a unique summary of the user input. Prompts are continuously added to the candidate list until the required number of candidates is reached.

\subsection{Seeds Generation and Selection}

Since LLMs are fundamentally based on text processing and generation, their core function is to understand and generate natural language text. However, many fuzz testing programs require inputs that are not purely text-based; for instance, libpng requires input in PNG format, and poppler requires input in PDF format. The solution proposed by ISC4DGF is to use a generation LLM to produce Python code that generates the required input format. The prompt used is: "Generate Python code so that the running result of the code is $\langle format \rangle$ file as a test case that may trigger $\langle CVE \rangle$ in $\langle PUT \rangle$." This approach results in an initial seed corpus in a valid input format compatible with the PUT.

ISC4DGF feeds candidate prompts into the generation LLM to create multiple Python scripts that may generate initial seeds capable of triggering the specified CVE. The sorting and selection of these candidate prompts depend on several factors, including the compilability of the script, the compliance of the seed type, and the size of the generated seeds. Since LLM-generated code is based on pattern recognition and large-scale training data, it may not always perfectly align with the specific environment or requirements. Several issues may arise with this generation method, potentially rendering the generated code unusable. Firstly, the generated code might contain syntax or logical errors because LLMs rely on language patterns without actual compilation or execution. Secondly, the code might lack necessary dependencies or library references, leading to execution failures even if the syntax is correct. Additionally, the code may not adhere to specific programming standards or formatting requirements.

To ensure that the generated code is functional, we manually run the LLM-generated code and select only those scripts that compile successfully and meet the format requirements of the PUT as the initial seeds. This process guarantees the validity and executability of the initial seeds.
Moreover, it is important to note that LLM-generated seed code may sometimes be too large, exceeding the maximum seed size that the fuzzer can handle. Oversized seeds not only affect the fuzzer's execution efficiency but may also prevent the fuzzer from processing these seeds correctly or even cause it to crash. Therefore, during the seed selection process, we screen out and exclude these oversized seeds to ensure that the fuzzer operates effectively while optimizing performance.

\section{Evaluation}

\begin{table*}
  \centering
  \caption{\textbf{Compared Fuzzers.} Selection of State-of-the-art Tools for Evaluation.}
  \begin{tabular}{ccc}
    \toprule
    \textbf{Fuzzer}&\textbf{Category}&\textbf{Description}\\
    \midrule
    \ AFLGo\cite{aflgo} & Directed &  Sophisticated seeds prioritization\\
    \ AFL\cite{afl} & Coverage &  Evolutionary mutation strategies\\
    \ Fairfuzz\cite{fairfuzz} & Coverage & Mutation with crucial bytes fixed\\
    \ Entropic\cite{entropic} & Coverage & Power scheduling with information entropy\\
  \bottomrule
\end{tabular}
\label{fuzzers}
\end{table*}

We implemented ISC4DGF, a directed greybox fuzzer that utilizes an initial seed corpus generated by LLMs. ISC4DGF is primarily implemented in Python and offers a lightweight alternative to traditional directed fuzzers like AFLGo \cite{aflgo}, which require significant manual effort to select and generate high-quality seeds that can trigger specified targets. ISC4DGF leverages GPT-4 as both the refinement LLM and generation LLM for prompt optimization and seed generation, given that this model represents the state-of-the-art in NLP-based reasoning tasks \cite{bubeck2023sparks}. Specifically, we use the gpt-4-0613 checkpoint provided by the OpenAI API \cite{gpt4}.
By default, we use AFL 2.57 \cite{afl} as the fuzzing engine. All experiments were conducted on a server running Ubuntu 18.04.6 LTS, equipped with two 64-core AMD EPYC 7763 CPUs @2.45GHz and 384GB of DDR4 RAM.

In this section, we evaluate the effectiveness of ISC4DGF by investigating the following research questions:

\textbf{RQ1:} How efficiently can ISC4DGF reproduce the vulnerabilities compared with other fuzzer?

\textbf{RQ2:} How does ISC4DGF improve the speed of triggering target vulnerabilities?

\textbf{RQ3:} Does ISC4DGF focus fuzzer coverage on target areas?

\textbf{Baseline.} We compare ISC4DGF with the fuzzers listed in Table \ref{fuzzers}. AFLGo\cite{aflgo} is the state-of-the-art directed gray-box fuzzer. We further choose AFL \cite{afl}, Fairfuzz\cite{fairfuzz} and Entropic \cite{entropic}, the fundamental greybox fuzzer and its updated versions, as the baseline.

\textbf{Benchmark.} To answer the research questions, we utilize the state-of-the-art benchmark, Magma \cite{magma}, which includes various CVEs from different programs commonly evaluated in advanced fuzzing studies \cite{shah2022mc2, srivastava2022one}.

\textbf{Configuration.} We followed the configuration recommended by the Magma \cite{magma} for our experiments. Each experiment was conducted over 10 trials with a 24-hour time limit. We applied the Mann-Whitney U Test \cite{mcknight2010mann} to demonstrate the statistical significance of ISC4DGF's contributions.

\subsection{Efficiency in Reproducing Vulnerabilities}

\begin{table*}
  \centering
  
  \caption{\textbf{Reproduction time for targets in Magma.} \textit{Time} indicates the average replication time (in seconds) over 10 runs. \textit{T:O.} indicates that the fuzzer could not replicate the target within the given time budget (24 hours). \textit{Ratio} and \textit{p} indicate the improvement ratio and p-value compared to ISC4DGF. }
  \resizebox{\textwidth}{!}{
  \begin{tabular}{cc|c|ccc|ccc|ccc|ccc}
    \toprule
    \multirow{2}{*}{\textbf{Bug ID}} & \multirow{2}{*}{\textbf{CVE}} & \multicolumn{1}{c|}{\textbf{ISC4DGF}} & \multicolumn{3}{c|}{\textbf{AFLGo}} & \multicolumn{3}{c|}{\textbf{AFL}} & \multicolumn{3}{c|}{\textbf{Fairfuzz}} & \multicolumn{3}{c}{\textbf{Entropic}}\\
    \cline{3-15}
    {}&{}&\textbf{Time}&\textbf{Time}&\textbf{Ratio}&\textbf{p}&\textbf{Time}&\textbf{Ratio}&\textbf{p}&\textbf{Time}&\textbf{Ratio}&\textbf{p}&\textbf{Time}&\textbf{Ratio}&\textbf{p}\\
    \midrule
    \ PNG001 & CVE-2018-13785 & 13884 & T.O. & N.A & - & T.O. & N.A & - & T.O. & N.A & - & T.O. & N.A & -\\
    \ PNG003 & CVE-2015-8472 & 12 & 14 & 1.17x & 0.02 & 15 & 1.25x & 0.01 & 3 & 0.25x & 0.03 & 4 & 0.33x & 0.01\\
    \ PNG007 & CVE-2013-6954 & 7070 & 28975 & 4.10x & \textless0.01 & T.O. & N.A & - & 7007 & 0.99x & 0.01 & 9967 & 1.41x & \textless0.01\\
    \ TIF007 & CVE-2016-10270 & 585 & 3761 & 6.43x & \textless0.01 & 6812 & 11.64x & \textless0.01 & 98 & 0.17x & 0.04 & 835 & 1.43x & 0.03\\
    \ TIF012 & CVE-2016-3658 & 3856 & 38764 & 10.05x & \textless0.01 & 14058 & 3.65x & \textless0.01 & 845 & 0.22x & 0.02 & T.O. & N.A & -\\
    \ TIF014 & CVE-2017-11613 & 6897 & 38480 & 5.58x & \textless0.01 & 18505 & 2.68x & 0.02 & 62319 & 9.04x & \textless0.01 & T.O. & N.A & -\\
    \ XML017 & CVE-2016-1762 & 15 & 16 & 1.07x & 0.02 & 18 & 1.2x & 0.01 & 20 & 1.33x & 0.03 & 14 & 0.93x & 0.02\\
    \ PDF010 & 	Bug \#10136 \cite{PDF010} & 175 & 37895 & 216.54x & 0.01 & 6794 & 38.82x & \textless0.01 & 37932 & 216.75x & \textless0.01 & 18790 & 107.37x & \textless0.01 \\
    \ PDF016 & CVE-2018-13988 & 32 & 187 & 5.84x & 0.01 & 265 & 8.28x & \textless0.01 & 13856 & 433x & \textless0.01 & 39 & 1.22x & 0.01\\
    \midrule
    \ \textbf{Avg.} &	- & - & \multicolumn{3}{c|}{\textbf{\textgreater 31.35x}} & \multicolumn{3}{c|}{\textbf{\textgreater 9.65x}} & \multicolumn{3}{c|}{\textbf{\textgreater 82.72x}} & \multicolumn{3}{c}{\textbf{\textgreater 18.78x}} \\
  \bottomrule
\end{tabular}
}
\label{RQ1}
\end{table*}

The main goal of ISC4DGF is to achieve directed fuzzing, specifically by efficiently reproducing known vulnerabilities. Directed fuzzing plays a critical role in vulnerability research, as it allows researchers to focus on specific areas of interest, increasing the likelihood of identifying critical issues in software systems. To evaluate ISC4DGF's performance in this regard, we conducted a comparative analysis with AFLGo\cite{aflgo}, a state-of-the-art directed fuzzer known for its precision and effectiveness. In our evaluation, we ran each fuzzer 10 times under identical conditions and measured the average time required to successfully reproduce the marked vulnerabilities in the Magma benchmark suite, which is widely recognized for its challenging and diverse set of real-world vulnerabilities. The results of this comparison are summarized in Table \ref{RQ1}, which lists the time required by each evaluated fuzzer to trigger different targets.

Compared to the directed fuzzer AFLGo, ISC4DGF demonstrated a marked enhancement, with a p-value of less than 0.05. On average, ISC4DGF outperformed AFLGo by a factor of 31.35x, reflecting a substantial enhancement in its directed fuzzing capabilities.
A key factor contributing to ISC4DGF's superior performance is its efficient approach to targeting specific vulnerabilities. AFLGo relies on additional static analysis to calculate distance metrics. According to our experiments, this process incurs considerable overhead, with each PUT requiring at least one hour of analysis time. In contrast, ISC4DGF achieves similar or better results with a much lower computational cost. While static analysis is crucial for AFLGo's targeting mechanism, it involves a significant time investment. This can become a bottleneck, especially in large-scale or time-sensitive testing scenarios.

\begin{table*}
  \centering
  \caption{\textbf{The number of reaches required to trigger the specified target. }If the target is not triggered within the allotted time, the number of times the fuzzer reaches the target within 24 hours is recorded.}
  \resizebox{\textwidth}{!}{
  \begin{tabular}{c|c|cc|cc|cc|cc}
    \toprule
    \multirow{2}{*}{\textbf{Bug ID}} & \multicolumn{1}{c|}{\textbf{ISC4DGF}} & \multicolumn{2}{c|}{\textbf{AFLGo}} & \multicolumn{2}{c|}{\textbf{AFL}} & \multicolumn{2}{c|}{\textbf{Fairfuzz}} & \multicolumn{2}{c}{\textbf{Entropic}}\\
    \cline{2-10}
    {}&\textbf{Reach}&\textbf{Reach}&\textbf{Ratio}&\textbf{Reach}&\textbf{Ratio}&\textbf{Reach}&\textbf{Ratio}&\textbf{Reach}&\textbf{Ratio}\\
    \midrule
    \ PNG001 & 279868596 & 109712322 & 0.39x & 28333100 & 0.10x & 138522498 & 0.49x & 303421286 & 1.08x\\
    \ PNG003 & 10 & 30901 & 3090.1x & 58190 & 5819x & 8648 & 864.8x & 20637 & 2063.7x\\
    \ PNG007 & 731047 & 16452364 & 22.51x & 10951093 & 14.98x & 6335404 & 8.67x & 1045254 & 1.43x\\
    \ TIF007 & 2 & 2582 & 1291x & 1880 & 940x & 401 & 200.5x & 4000 & 2000x\\
    \ TIF012 & 603697 & 32891737 & 54.48x & 28300878 & 46.88x & 2730622 & 4.52x & 3903720 & 6.47x\\
    \ TIF014 & 203452 & 514837 & 2.53x & 363487 & 1.77x & 2141687 & 10.53x & 121037 & 0.59x\\
    \ XML017 & 84 & 2339 & 27.85x & 2503 & 29.80x & 2501 & 29.77x & 517 & 6.15x\\
    \ PDF010 & 7027 & 37895 & 5.39x & 147850 & 21.04x & 1050 & 0.15x & 3655 & 0.52x\\
    \ PDF016 & 2010 & 234893 & 116.86x & 143734 & 71.51x & 10869658 & 5407.79x & 32252 & 16.05x\\
    \midrule
    \ \textbf{Avg.} & - & \multicolumn{2}{c|}{\textbf{512.35x}} & \multicolumn{2}{c|}{\textbf{771.68x}} & \multicolumn{2}{c|}{\textbf{725.25x}} & \multicolumn{2}{c}{\textbf{455.11x}} \\
    
  \bottomrule
\end{tabular}
}
\label{RQ2}
\end{table*}

In contrast, ISC4DGF leverages a more streamlined process, where the time cost associated with generating and refining prompts is negligible. This efficiency not only reduces the overall testing time but also allows for more iterations and broader exploration within the same timeframe, thereby increasing the likelihood of uncovering vulnerabilities.
This comparison highlights how the design of the initial seed corpus can effectively enhance directed fuzzing capabilities, enabling ISC4DGF, which is built on the coverage-guided AFL, to outperform even established directed fuzzers like AFLGo.

Compared to non-directed fuzzers, specifically coverage-guided greybox fuzzers, ISC4DGF demonstrated a clear advantage by detecting 6 additional bugs and achieving an average speed improvement of over 37.05 times. This performance boost underscores the effectiveness of ISC4DGF in generating high-quality initial seeds that guide the fuzzing process more efficiently.
While ISC4DGF's performance may be less pronounced in certain cases, such as Fairfuzz's PNG003, TIF007, and Entropic's PNG003 and XML017, it's important to note that the time to trigger these targets was relatively short (less than 1000 seconds). We consider these results acceptable within the context of randomized fuzzing scenarios.
On targets requiring deeper exploration, such as TIF014 and PDF010, ISC4DGF's performance was notably superior. In these cases, where traditional fuzzers might struggle due to the need for extensive exploration to reach deeply buried code segments, ISC4DGF's ability leads to better results due to efficiently guide the fuzzing process and focus on critical areas.

Overall, ISC4DGF demonstrated an average performance improvement of 35.63 times compared to existing popular fuzzers on the Magma dataset, detecting 7 additional bugs. Notably, ISC4DGF was the only fuzzer capable of triggering PNG001 within 24 hours, confirming that it not only enhances the efficiency of vulnerability detection but also has the ability to uncover vulnerabilities that other fuzzers might miss.

\subsection{Fast-tracking Exploits}

In fuzz testing, reaching a target does not guarantee that the target vulnerability will be successfully triggered. A fuzzer may reach the target code segment multiple times, but if the input or state does not satisfy the conditions needed to trigger the vulnerability, the issue will remain unexposed. In such cases, while the fuzzer's path exploration capabilities are validated, its actual efficiency in detecting vulnerabilities may be limited.
To address this, we analyzed the number of times different fuzzers reached the designated target before successfully triggering the vulnerability. This analysis allowed us to assess whether ISC4DGF's initial seed set could trigger target vulnerabilities more quickly with fewer target reaches. If a fuzzer failed to trigger the target vulnerability within the specified time, we recorded the number of times it reached the target after 24 hours of execution. The results of our analysis are summarized in Table \ref{RQ2}.

\begin{figure*}
    \centering
    \subfigure[Coverage Comparison]{
        \includegraphics[width=0.45\textwidth]{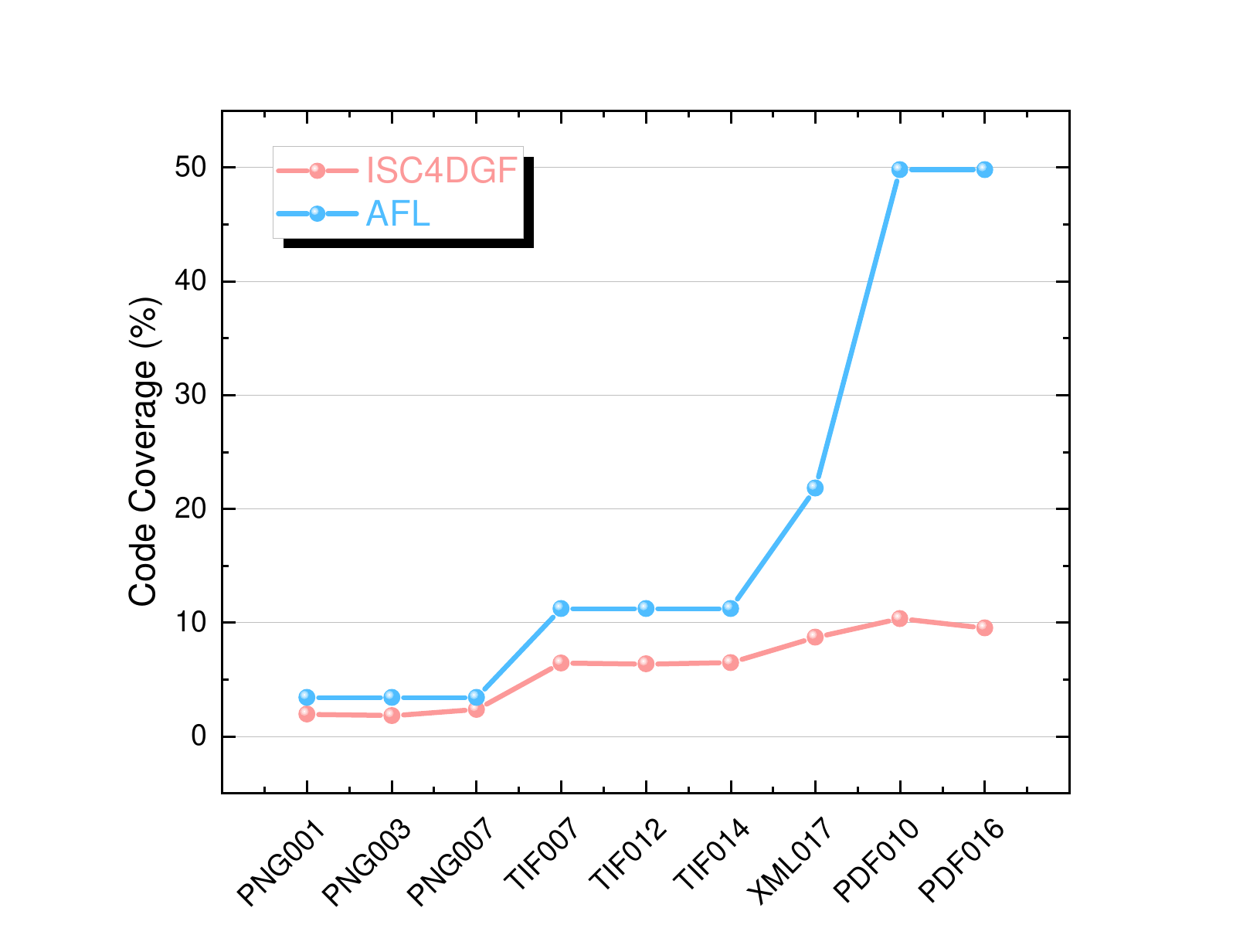}
        \label{RQ3a}
    }
    \hfill
    \subfigure[Reach Comparison]{
        \includegraphics[width=0.45\textwidth]{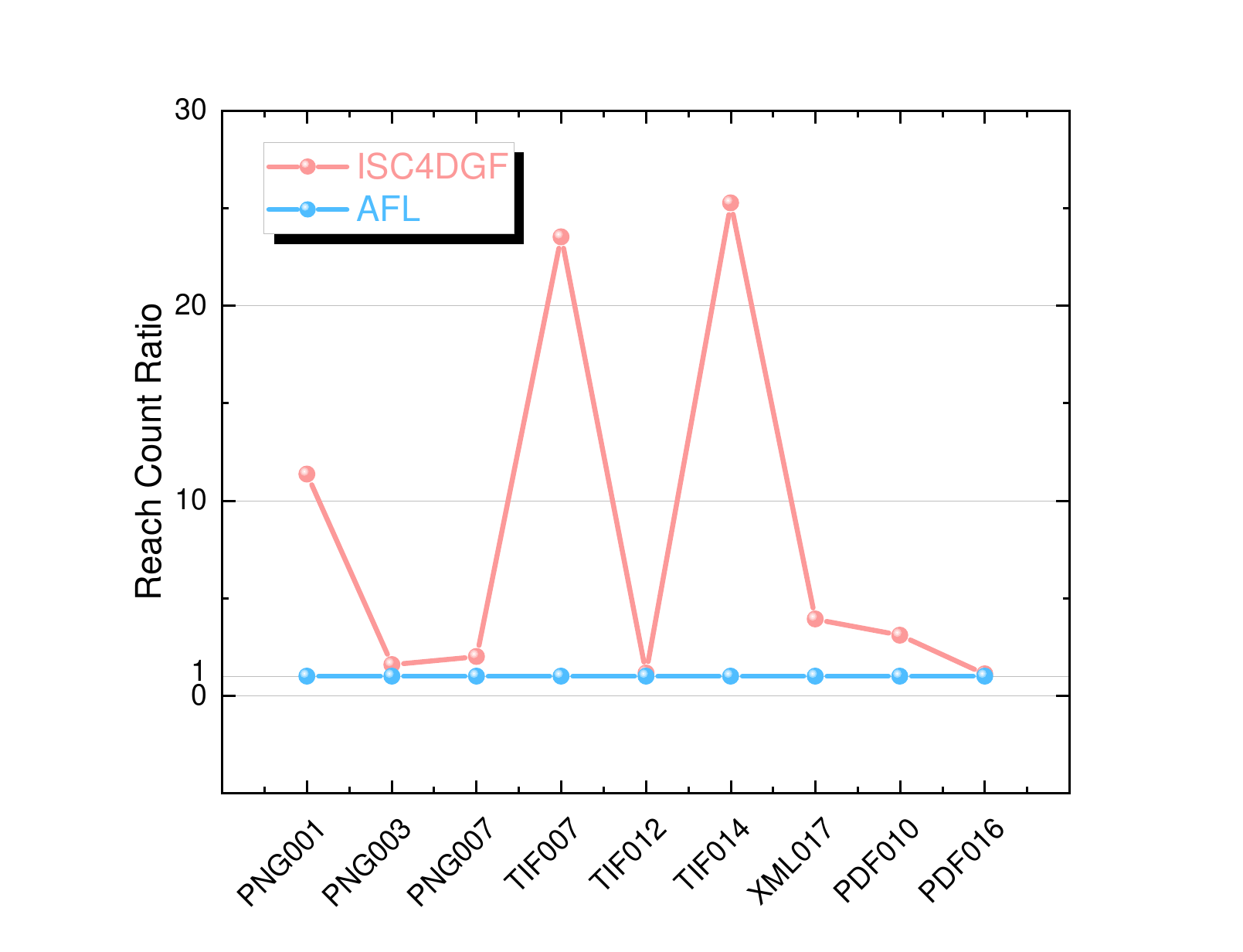}
        \label{RQ3b}
    }
    \caption{\textbf{Comparison of Code Coverage and Target Reach Efficiency. }\textbf{Figure (a)} illustrates the code coverage, while \textbf{Figure (b)} compares the number of target reaches, both for ISC4DGF and AFL.}
    \label{RQ3}

\end{figure*}

ISC4DGF demonstrated a significant reduction in the number of target reaches needed to trigger vulnerabilities, achieving up to 771.68x fewer target reaches compared to other fuzzers. This efficiency underscores ISC4DGF's ability to trigger specified targets more quickly and with greater precision, validating its directed detection capabilities.
To further understand ISC4DGF's performance, we conducted an in-depth analysis of three specific experiments where it excelled: PNG003, TIF007, and XML014. In these cases, ISC4DGF managed to trigger the specified targets with fewer than 100 target reaches. Compared to other fuzzers, ISC4DGF reduced the number of target reaches by factors of 2959.4x, 1107.88x, and 23.39x, respectively. This substantial reduction highlights the effectiveness of our initial seed set, which not only guides the fuzzing process more efficiently but also contains the necessary semantic information to trigger the specified vulnerabilities. The ability of ISC4DGF to focus on relevant code paths and quickly achieve the desired outcomes clearly demonstrates its enhanced directed fuzzing capabilities.
Additionally, the experiment involving PNG001 provided further insights into ISC4DGF's effectiveness in challenging scenarios. In this case, ISC4DGF registered a significantly higher number of target reaches compared to other fuzzers because the others were unable to trigger the target vulnerability within the 24-hour testing period. In contrast, ISC4DGF persisted with a substantial number of target reaches and ultimately succeeded in triggering the vulnerability within the limited time.

\subsection{Coverage Focus on Target Areas}

ISC4DGF is built on the AFL, which is renowned for its ability to enhance vulnerability detection through coverage-guided fuzzing. AFL's approach emphasizes maximizing code coverage, enabling the fuzzer to extensively explore the PUT, thereby increasing the likelihood of discovering unknown vulnerabilities across the software.
However, ISC4DGF diverges from AFL's philosophy by focusing less on achieving maximum code coverage and more on swiftly triggering specific targets. ISC4DGF is effective in scenarios involving known or suspected vulnerabilities, where quickly reaching the relevant code areas is more critical than exploring the entire codebase.
To assess whether ISC4DGF prioritizes targeting over code coverage, we compared the code coverage and the number of target reaches achieved by ISC4DGF and AFL within a 24-hour period. Figure \ref{RQ3} presents our experimental results.

In Figure \ref{RQ3a}, we compared the code coverage achieved by ISC4DGF and AFL across various targets. As expected, ISC4DGF demonstrated lower code coverage compared to AFL, with an average reduction of 12.38\%.
However, this reduction in coverage is more than offset by ISC4DGF's performance, as it achieved a 8.12x increase in target reach counts compared to AFL, as shown in Figure \ref{RQ3b}. This result suggests that ISC4DGF effectively concentrates its fuzzing efforts on the most relevant sections of the code, dedicating more resources to exploring these critical areas rather than spreading its efforts across the entire codebase.
By prioritizing target hits over broader code coverage, ISC4DGF can trigger vulnerabilities more quickly.

\subsection{Discussion}

\textbf{Difficulties in Detecting Hard-to-find Bugs.}
While ISC4DGF primarily improves the speed of vulnerability detection, it can struggle to identify bugs that other fuzzers might miss. This limitation suggests that ISC4DGF's enhancements are more focused on efficiency rather than expanding the scope of detected vulnerabilities.
The increased speed is a notable advantage, especially in scenarios where time efficiency is critical. However, while ISC4DGF excels in rapidly detecting vulnerabilities, it may not always enhance the discovery of new or previously undetected bugs.
Future research should explore combining ISC4DGF's speed benefits with other techniques that improve the detection of hard-to-find vulnerabilities.

\textbf{Seed Mutation in Directed Fuzzing.}
ISC4DGF's current advancements have primarily focused on optimizing the initial seed corpus, which has enhanced the fuzzer's directed capabilities, enabling it to target specific vulnerabilities more efficiently from the start. 
Other steps in the fuzzing process, particularly the seed mutation phase, still rely on traditional techniques that may not fully capitalize on the initial advantages provided by ISC4DGF.
Future research could explore integrating LLMs into the seed mutation phase. By leveraging LLMs to generate higher-quality seeds, it may be possible to further enhance the fuzzer's directed capabilities, leading to more effective vulnerability detection.

\section{Related Work}

\textbf{Directed Grey-box Fuzzing.} Many works have been proposed to improve directed fuzzing efficiency.Existing work can be broadly categorized into three types. The first category is distance-minimization-based approaches, such as AFLGo\cite{aflgo} and Hawkeye\cite{hawkeye}. These works primarily focus on using heuristic metrics to guide the fuzzer closer to the target, enabling faster and more frequent triggering of the target vulnerabilities. The second category is input-reachability-based approaches, exemplified by tools like BEACON\cite{beacon} and FuzzGuard\cite{fuzzguard}, which focus on the reachability of seeds to specific targets. By minimizing the impact of irrelevant inputs during testing, these works enhance the directed fuzzing capabilities. The third category is sequence-guided approaches, such as Berry\cite{berry} and UAFL\cite{uafl}, which address the impact of execution order on triggering target vulnerabilities. These works analyze the execution sequences leading to target crashes and guide the fuzzer to execute these sequences in order to trigger the crashes.

\textbf{Large Language Models Based Fuzzing.}The integration of LLMs has introduced significant advancements in fuzzing, enabling more sophisticated mutation operations that maintain input validity. LLMs enhance fuzzing by understanding and modifying data types in context, allowing for complex changes such as rewriting code fragments and altering data structures.
Recent approaches like Fuzz4All\cite{xia2024fuzz4all} and InputBlaster \cite{liu2024testing}have innovatively combined prompt engineering with seed mutation to improve fuzzing effectiveness. Fuzz4All \cite{xia2024fuzz4all} refines complex user inputs into concise prompts, which are then used to generate diverse code snippets for fuzz testing, with each iteration updating prompts to avoid redundancy. InputBlaster \cite{liu2024testing}, on the other hand, uses contextual information from mobile app interfaces as prompts to generate specialized text inputs and inferred constraints, guiding subsequent mutations.
Furthermore, ParaFuzz \cite{yan2024parafuzz} introduced a Meta prompt method to perform semantically consistent statement mutations using ChatGPT, while TitanFuzz leveraged models like CodeX to generate varied input data through strategic masking and mutation operations.

\section{Conclusion}

In this paper, we presented ISC4DGF, a novel approach to improving directed grey-box fuzzing by generating optimized initial seed corpora using Large Language Models. By focusing on the design of the initial seed corpus, ISC4DGF enhances the efficiency and precision of fuzzing, leading to faster and more accurate detection of specific vulnerabilities. Our experiments demonstrated that ISC4DGF significantly outperforms state-of-the-art fuzzers, achieving a 35.63x speedup and 616.10x fewer target reaches. These results underscore the effectiveness of LLMs in improving directed fuzzing, particularly when precise and efficient vulnerability detection is needed.

\bibliographystyle{plain}
\bibliography{ref.bib}

\label{last-page}
\end{multicols}
\label{last-page}
\end{document}